\documentclass{article}

\usepackage{PRIMEarxiv}

\usepackage[utf8]{inputenc} % allow utf-8 input
\usepackage[T1]{fontenc}    % use 8-bit T1 fonts
\usepackage{hyperref}       % hyperlinks
\usepackage{url}            % simple URL typesetting
\usepackage{booktabs}       % professional-quality tables
\usepackage{amsfonts}       % blackboard math symbols
\usepackage{nicefrac}       % compact symbols for 1/2, etc.
\usepackage{microtype}      % microtypography
\usepackage{lipsum}
\usepackage{fancyhdr}       % header
\usepackage{graphicx}       % graphics
\usepackage{threeparttable}
\usepackage{multirow}
\usepackage{array}
\graphicspath{{media/}}     % organize your images and other figures under media/ folder

\title{scpQCA: Enhancing mvQCA Applications through Set-Covering-Based QCA Method
}

\author{
  Manqing Fu \\
  School of Big Data \\
  Fudan University \\
  Shanghai\\
  \texttt{mqfu23@m.fudan.edu.cn} \\
  \texttt{fumanqing@outlook.com} \\
}

\begin{document}
\maketitle

\begin{abstract}
In fields such as sociology, political science, public administration, and business management, particularly in the direction of international relations, Qualitative Comparative Analysis (QCA) has been widely adopted as a research method. This article addresses the limitations of the QCA method in its application, specifically in terms of low coverage, factor limitations, and value limitations. scpQCA enhances the coverage of results and expands the tolerance of the QCA method for multi-factor and multi-valued analyses by maintaining the consistency threshold. To validate these capabilities, we conducted experiments on both random data and specific case datasets, utilizing different approaches of CCM (Configurational Comparative Methods) such as scpQCA, CNA, and QCApro, and presented the different results. In addition, the robustness of scpQCA has been examined from the perspectives of internal and external across different case datasets, thereby demonstrating its extensive applicability and advantages over existing QCA algorithms.
\end{abstract}

% keywords can be removed

\section{Introduction}
Qualitative Comparative Analysis (QCA) is a widely applied research method developed by Ragin in 1987 \cite{marx_origins_2014} and subsequent scholars; that is combined with set theory to analyze the necessary and sufficient conditions for the outcome \cite{greckhamer_studying_2018}. Initially, QCA was used to deal with strictly dichotomous dataset (csQCA), and later expanded to continuous data (fsQCA) and multivalued sets (mvQCA). However, in COMPASSS (COMPArative methods for Systematic cross-caSe analySis), there are a lot of discussions about fsQCA data needing additional binary calibration and seldom literatures using mvQCA \cite{liu_search_2021}. The lack of application of mvQCA problems on one hand is caused by the short of corresponding introductory textbooks \cite{haesebrouck_added_2016}, on the other hand, the difficulty of dataset construction and the undeveloped algorithm. scpQCA algorithm introduced in this paper is to break through the limitations of the latter.

scpQCA is committed to improving the following three specific problems on the basis of the existing QCA algorithm: low solution coverage, difficulty in running multi-factors datasets, and difficulty in running multi-value datasets. Through our research, the above problems are mainly caused by the using the Quine-McCluskey (QMC) algorithm in traditional QCA software \cite{whitaker_coincidence_2020}. Therefore, the main improvement of the scpQCA method is to replace the Quine-McCluskey algorithm module with the set covering problem algorithm module. Beyond this, scpQCA retains most of the processing steps of the QCA algorithm and related parameter requirements. The retaining steps are, for example, necessity analysis, truth table analysis, sufficient simplification. In addition, necessary and sufficiency consistency threshold and the cutoff number of cases are retained, while we add a unique cover parameter to help the calculation.

Quine-McCluskey algorithm should be replaced because of its proven flaws. For example, a dataset with many factors will lead the Quine-McCluskey algorithm to an exponential increase in spatial complexity and runtime \cite{jain_optimization_2008}, the inherent instability of heuristics will lead the Quine-McCluskey algorithm to \cite{jadhav_modified_2012}, and so on. scpQCA is simpler and more transparent than using qmcQCA algorithm. For the former, scpQCA only outputs the parsimonious solution to avoid the choice confusion caused by complex and intermediate solutions \cite{haesebrouck_introduction_2021, oana_qualitative_2021}; For the latter, scpQCA can be stably reproduced with defined parameter settings, avoiding interference from different calibration scales and other intermediate procedures \cite{ide_qca_2022}. In addition, the advantage of the SCP algorithm over the QMC algorithm is also the scalability of the algorithm, and its powerful IEEE application background can be iterated and upgraded faster according to the update of requirements.

This article is divided into four sections. Section \ref{sc1} introduces the components and operational steps of the scpQCA method. Section \ref{sc2} brings in the requirements of improving the mvQCA in qualitative and quantitative analysis usage. Section \ref{sc3} discusses how the scpQCA method has improved the existing QCA methods from high coverage situation, multi-factor situation and multi-value situation. Section \ref{sc4} demonstrates the superiority of the scpQCA method over existing methods through three specific case studies. Section \ref{sc5} tests the robustness of the scpQCA method from internal validity and external validity. 

\section{scpQCA Method{\label{sc1}}}
As a new method of Qualitative Comparative Analysis (QCA), scpQCA has two unique advantages. Firstly, its operation and results type are both simple. scpQCA only produces parsimonious results through a rigorous simplification procedure, thereby avoiding the misunderstanding associated with different types of solutions \cite{haesebrouck_introduction_2021, oana_qualitative_2021}. Secondly, its computational process is transparent. Other QCA software results are easily affected by different calibration standards or intermediate steps \cite{ide_qca_2022}. As for scpQCA, all procedures and intermediate results (e.g., necessary conditions, candidate rule lists) are preserved and traceable.

\subsection{Necessary conditions}
Necessity analysis holds a very important position in Ragin's QCA (Qualitative Comparative Analysis) theory, yet most QCA results tend to neglect it (e.g., \cite{haesebrouck_alternative_2019}). scpQCA emphasizes necessity analysis for two main reasons: firstly, to fulfill the prerequisite of QCA as a set-theoretic causal deduction, and secondly, to differentiate scpQCA method from another CCM (Configurational Comparative Methods) approach, CNA (Coincidence Analysis) \cite{baumgartner_causal_2020}. Although both methods can achieve comparable levels in terms of consistency and coverage metrics, they cannot be substituted for one another due to their different theoretical underpinnings.

The method used by scpQCA to determine whether a condition is a necessary condition involves calculating their consistency. Taking the calculation of consistency in mvQCA as an example, as described by \cite{thiem_parameters_2015,thiem_navigating_2014}:

\begin{equation}
  Consistency_N(C=c\rightarrow O=o)=\frac{\sum_{i=1}^{n}\mathbb{I} _{(C_i=c, O_i=o)}}{\sum_{i=1}^{n}\mathbb{I}_{(O_i=o)}}
\end{equation}

Conventionally, $Consistency_N>0.9$ is considered as a necessary condition for the outcome. After the simplification steps of sufficiency, it will be conjoined with the disjunctive form of the configuration. The result obtained in this way is the necessary and sufficient condition that satisfies the logical relationship.

\subsection{Candidate Rules}
In the process of sufficiency analysis in QCA, it is necessary to first filter out the sufficient conditions that meet the consistency standard. These sufficient conditions are typically composed of the conjunction of conditions. Conventionally, the conjunction of factors is considered as the sufficient conditions for the solution. Sufficient consistency and necessary consistency have opposite mapping directions on sets, and the calculation formula is as follows \cite{thiem_navigating_2014}:
\begin{equation}
  Consistency_S(C=c\rightarrow O=o)=\frac{\sum_{i=1}^{n}\mathbb{I} _{(C_i=c, O_i=o)}}{\sum_{i=1}^{n}\mathbb{I}_{(C_i=c)}}
\end{equation}
To obtain the disjunction of these conjunctions, qmcQCA first uses a truth table to sort out the case coding.

Here we use the `Openness and Innovation' dataset \cite{schneider_two-step_2019} to show how the candidate rules runs in scpQCA. Table \ref{tb1} displays a portion of the truth table obtained from the original research (Schneider 2018). Table \ref{tb2} shows the candidate rule list derived from the same dataset using scpQCA\footnote{Take the rows containing $LE=1$ and $LE=1$ away, for they satisfy the necessary condition of $consistency_N$.}.

\begin{table}
  \caption{Truth table of second step all remote conditions\label{tb1}}
  \centering
  \begin{tabular}{cccccccccc}
    \toprule
    \multicolumn{7}{c}{Condition}                   \\
    \cmidrule(r){2-8}
    Row     & MS & MC & PI & LE & LP & PV & ED\textsuperscript{1} & Consist. fuzzy & Cases \\
    \midrule
    I & 0  & 1 & 1 & 1 & 1 & 0 & 1 & 0.93 & FR, PL, PT \\
    II & 0 & 1 & 0 & 1 & 1 & 1 & 1 & 0.89 & ES  \\
    III & 0 & 1 & 0 & 1 & 1 & 0 & 1 & 0.85 & BE, GR  \\
    IV & 0 & 0 & 1 & 1 & 1 & 1 & 1 & 0.74 & FI  \\
    V & 0 & 1 & 0 & 1 & 0 & 0 & 1 & 0.67 & IT  \\
    \bottomrule
  \end{tabular}
   \begin{tablenotes}
     \item[1] \textsuperscript{1}MS: military stretch, MC: military capabilities, PI: peacekeeping involvement, LE: left executive, LP: left parliament; PV: parliamentary veto; ED: electoral distance
   \end{tablenotes}
\end{table}

\begin{table}
  \caption{Candidate rule list of second step all remote conditions by scpQCA\label{tb2}}
  \centering
  \begin{tabular}{ccccccccccc}
    \toprule
    \multicolumn{7}{c}{Condition}                   \\
    \cmidrule(r){2-8}
    Row     & MS & MC & PI & LE & LP & PV & ED & Consist. crispy\textsuperscript{1} & Cases & Table \ref{tb1} row\\
    \midrule
    1 & 0 & - & - & - & - & 0 & 1 & 0.8333 & BE, FR, GR, IT, PL, PT, (SI) &{}\\
    3 & 0 & - & 1 & - & 1 & - & - & 1.0 & FI, FR, PL, PT & I, IV\\
    9 & 0 & - & 1 & - & - & - & - & 0.8 & FI, FR, PL, PT, (CZ) &{}\\
    10 & 0 & - & - & - & 1 & 0 & - & 0.8333 & BE, FR, GR, PL, PT, (SI) &{}\\
    17 & - & 1 & - & - & - & 0 & - & 0.8571 & BE, FR, GR, IT, PL, PT, (GB) &{}\\
    20 & 0 & 1 & - & - & - & 0 & - & 1.0 & BE, FR, GR, IT, PL, PT & I, III, V\\
    27 & 0 & 1 & - & - & 1 & - & - & 1.0 & BE, ES, FR, GR, PL, PT &{}\\
    37 & - & 1 & - & - & 1 & 0 & - & 1.0 & BE, FR, GR, PL, PT &{}\\
    40 & - & 1 & - & - & 1 & - & - & 1.0 & BE, ES, FR, GR, PL, PT & I, II, III\\
    \bottomrule
  \end{tabular}
   \begin{tablenotes}
     \item[1] \textsuperscript{1}Dichotomous data using $candidate\_rules$ function with $decision\_label=1$, $consistency=0.8$, $cutoff=4$.
   \end{tablenotes}
\end{table}

Through Table \ref{tb2}, it can be seen that the main difference between candidate rules and the truth table is that candidate rules do not have to include all conditions. As a result, scpQCA is a top-down approach that first filters the simplified conjunctions (also known as atomic minus-formulas in CNA), and then performs a maximum coverage set operation on these conjunctions. Additionally, during the calculation of candidate rules, a screening is conducted on the $consistency_S$ of these conjunctions, which makes the solution consistency of scpQCA robust.

Just as mentioned previously, CNA and scpQCA differ in their approaches to the deduction of complex causal pathways. Firstly, CNA aims to identify all potential sufficient configurations, which may lead the result disjunction in containing a larger number of atomic minus-formulas under given conditions. This creates more challenges for the users to interpret than explaining the distinctions between the most parsimonious, intermediate, and complex solutions in general QCA. Additionally, the combinations of atomic minus-formulas are based on complexity criteria. Therefore, it's rare for a factor's different values to play a role in different atomic minus-formulas. It can't be denied that the subsequent simplification methods of CNA provide a new perspective on causal mechanisms. Furthermore, the consistency and coverage of the highest-ranked configuration solution are comparable to those of scpQCA within a certain range, as demonstrated in Section \ref{sc2}.

Returning to the QCA methodology itself, although the approaches of scpQCA and qmcQCA differ in their simplification processing logic, there are commonalities in their results. Rows 3, 20, and 40 in Table \ref{tb2} represent combinations of the final scpQCA solution, and thus, the configurations of these three rows will be the focus of the main discussion. Row 3 in Table \ref{tb2} represents a candidate rule for $ms*PI*LP$, with a consistency of 1.0 and covering four cases; in contrast, in Table \ref{tb1}, these four cases are respectively part of Rows I and IV. Therefore, Row 4 in the candidate rule list can be considered a simplification of Rows I and IV in the truth table. Row 20 in Table \ref{tb2} represents $ms*MC*pv$, with a consistency of 1.0 and covering six cases; in Table \ref{tb1}, these six cases are respectively part of Rows I, III, and V, which means that Row 20 in the candidate rule list is a simplification of Rows I, III, and V in the truth table. Row 40 in Table \ref{tb2} represents $MC*LP$, with a consistency of 1.0 and covering six cases; in Table \ref{tb1}, these six cases are respectively part of Rows I, II, and III, indicating that Row 40 in the candidate rule list is a simplification of Rows I, II, and III in the truth table.

Besides, the fundamental difference between these two QCA methodology is clear. The scpQCA method, which limits the number of unique covering cases, yields configurations that are more independen, thereby providing a more representative interpretation. Conversely, qmcQCA, in an effort to minimize computational space complexity, often merges structurally similar rows from the truth table before simplification operation. As can be seen in the 'cases' columns of Table \ref{tb1} and \ref{tb2}, this approach can results in some events being covered by multiple configurations, while some events being missed and remain uncovered in the qmcQCA results.

\subsection{Simplification}
\begin{figure}
  \centering
  % \fbox{\rule[-.5cm]{4cm}{4cm} \rule[-.5cm]{4cm}{0cm}}
  \caption{Logical structure of scpQCA.\label{fg1}}
  \includegraphics[width=0.75\textwidth]{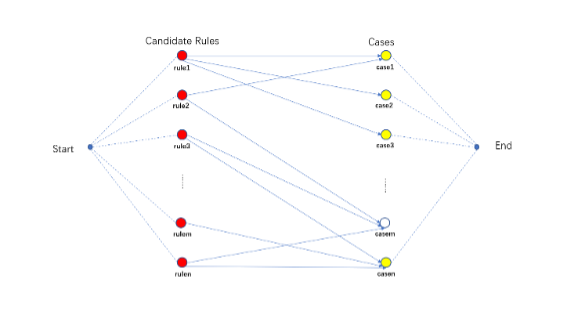}
\end{figure}

The set covering problem is a kind of solution method for operation research that is extremely widely applied, such as transportation networks for the flow of people, flow of vehicles, and flow of goods \cite{farahani_covering_2012}. Schilling et al.~\cite{schilling_review_1993} categorized set covering models into two types: the set covering problem that mandates a minimum level of coverage and seeks to optimize maximal coverage, and the maximal covering location problem that aims to optimize location selection while maximizing coverage. The simplification process in CNA is more akin to the former, while scpQCA is of the latter type. Regarding optimization algorithms, Caprara, et al. \cite{caprara_algorithms_2000} divided set covering problems into two categories: the Linear Programming (LP) relaxation and the heuristic algorithm. The LP method is more suitable for continuous variables, whereas heuristic algorithms have an advantage with integer and finite discrete sets. scpQCA was specifically developed for mvQCA-type data, hence it employs the greedy heuristic algorithm. The logical structure of scpQCA's greedy algorithm is shown in Figure \ref{fg1}.

In Figure \ref{fg1}, red nodes represent candidate rules, yellow nodes represent cases where the outcome is the focus of the study, and white circles represent cases where the outcome pertains to the others. The arrows from candidate rules column to cases column indicate the coverage relationship between the rule and the case. Candidate rules may cover either yellow nodes or white nodes, but the consistency of the yellow nodes in all covered events must always exceed a pre-set consistency threshold. The start node is connected to all candidate rule nodes, while the end node is exclusively connected to yellow case nodes.

The principle of the scpQCA simplification process is, in simple terms, to utilize a greedy algorithm to identify the minimal combination of candidate rules from the `Candidate Rules' column that can cover the maximum number of yellow nodes. Additionally, to ensure that the configurations in the solution are non-redundant, the simplification process incorporates a unique cover parameter to guarantee that each selected candidate rule has its own case characteristics. According to the criteria mentioned by \cite{ragin_measurement_2008}, the empirical value of the unique cover is typically an integer no less than 2, increasing with the total number of cases. The greedy algorithm has proven to be the most effective heuristic method tested to date. The scpQCA package has been uploaded to PyPI\footnote{\url{https://pypi.org/project/scpQCA}} and GitHub\footnote{\url{https://github.com/Kim-Q/scpQCA}} and the related supplement of this work can be found in Harvard Dataverse\footnote{\url{https://dataverse.harvard.edu/dataset.xhtml?persistentId=doi:10.7910/DVN/34MOHT&version=DRAFT}}.

The solution of scpQCA is a logical conjunction of the necessary condition and the simplification of sufficient conditions, meaning that the solution constitutes the necessary and sufficient conditions for the outcome. In the field of logic, there exists an `if and only if' relationship between the solution and the outcome; in causal mechanism, there is a close causal relationship between the solution and the outcome.

\section{Challenges{\label{sc2}}}
Charles Ragin (2014) posited that the majority of social facts, particularly behaviors and social outcomes, are driven by the nonlinear and even multidirectional interactions of multiple factors. However, like statistical analysis, \cite{ragin_comparative_2014} understanding is still based on a consideration of factors, without taking into account the critical role of mechanisms\footnote{The vast majority of regression analyses, including those by Pearl and Mackenzie \cite{pearl_book_2018}, have an incorrect understanding of `mechanism'. A mechanism is a process. It is a pathway, but not merely a pathway composed of factors alone.} \cite{bunge_mechanism_1997}. Therefore, there is no `sufficient condition' that is entirely composed of factors and their combinations; factors can be, at most, `necessary conditions'. In fact, it is only the combination of factors and mechanisms that can constitute a `necessary and sufficient condition'. Utilizing the concept of necessary and sufficient conditions from set theory as the basis for causal inference is a theoretical cornerstone of the QCA method, as adhered to in the development process of our scpQCA.

The widespread application of mvQCA is crucial for the QCA method, as it is considered the most qualitatively insightful among the three types of QCA analyses. The qualitative essence of mvQCA stems from a detailed explanation of multi-value calibration, which is lacking in csQCA and not available in fsQCA \cite{liu_search_2021}. Calibration refers to the transformation of abstract variables into values with interpretive meaning \cite{ragin_fuzzy_2009}. Its scope includes not only the conversion of fuzzy data into finite degree values \cite{mendel_new_2018} but also the assignment of values to multi-value data. scpQCA oppose a lazy approach to calibration, such as calibrating fuzzy or multi-value datasets into binary based solely on a fixed formula from direct or indirect methods \cite{ragin_measurement_2008}. In fact, calibrating the original sources into multi-value format and conducting analysis is a more appropriate application of an expert QCA user's knowledge.

Furthermore, the QCA method should not only be cautious with the interpretive content of calibration but also broadly collect cases related to the research question and regard high coverage as the standard for evaluating a good solution. Firstly, as a set-theoretic method, if cases are selectively sought with bias, the conclusions will inevitably be subjective as well. To ensure scientific rigor and the fairness of quantitative analysis, it is essential to collect and calibrate relevant positive and negative cases as much as possible. As the collection of cases expands, the QCA solution will more closely approximate the actual results. Secondly, the emphasis on coverage depends on the level of trust in the case dataset. If the calibration of all cases is accurate, a solution with higher coverage is invariably more credible than one with lower coverage at an approximate consistency level. Utilizing Karnaugh map is a method for simplifying mvQCA \cite{rushdi_utilization_2018}. Although using K-map algorithm requires a higher spatial complexity, it is coincidence with our understanding of improving the coverage of QCA results in relation to such set-theoretic problems.

Although both scpQCA and CNA \cite{baumgartner_causal_2020,whitaker_coincidence_2020} are top-down methods and share many of the same advantages comparing with qmcQCA, scpQCA still cannot be replaced by CNA. The most significant difference between the two lies in the fact that CNA does not conduct an analysis of necessary and sufficient conditions.  In the sufficiency analysis phase, both CNA and scpQCA begin by filtering for atomic minus-formulas with higher consistency, which yields a more stable solution consistency compared to qmcQCA. In the simplification phase, CNA arranges these atomic minus-formulas into a series of conjunctions of configurations based on the set complexity, and ranks them from highest to lowest solution coverage; in contrast, scpQCA employs a set covering problem algorithm to identify the unique cover and most typical combination of causal pathways, with solution coverage being one of the important selection criteria. According to subsequent experimental comparisons, under normal circumstances, the solution coverage of both scpQCA and CNA tends to be higher than that of qmcQCA.

In the forthcoming discussion, we will delve into an in-depth examination of three prevalent issues that are currently being encountered within the realm of QCA. To evaluate the performance of scpQCA, CNA, and QCA on high coverage, multi-factor, and multi-value problems, random datasets with specified causal mappings are constructed in the following experiments. Thiem~\cite{thiem_introducing_2012} mentioned that, up till 2012, there were typically six products in the QCA software market: Tosmana, fs/QCA, KirqST, fuzzy, QCA3, and QCA. The QCA method `QCApro', an enhanced algorithm package on R platform, is the focus of this section's QCA comparison.

\subsection{High Coverage}
Rubinson \cite{rubinson_presenting_2019} has previously highlighted the issue of low solution coverage in QCA algorithms; De Meur et al. \cite{de_meur_addressing_2009} addressed the situation of varying coverage ranges among results by introducing the concept of `case sensitivity'—adding unobserved cases to the truth table analysis—to differentiate outcomes with lower coverage rates. However, this approach was later refuted by Thiem \cite{thiem_beyond_2022}. As a heuristic algorithm, the Quine-McCluskey algorithm  cannot guarantee the coverage of solutions \cite{safaei_quine-mccluskey_2007}. When computational capacity is limited, QCA may compensate by economizing on the computational space required for simplification, which is also a reason why QCA can produce some lengthy solutions that are difficult to analyze.

As previously mentioned, the emphasis on coverage depends on the level of trust in the cases. In development, it has been discovered that the mathematical foundation of QCA is set theory and Boolean algebra \cite{ragin_comparative_2014,de_meur_addressing_2009}, which allows QCA to connect with basic logic studies, such as necessary conditions, sufficient conditions, necessary and sufficient conditions, INUS conditions \cite{mackie_causes_1965}, and SUIN conditions \cite{mahoney_logic_2009}. QCA methodologies can also be improved through methods from computer science and combinatorial mathematics. The concept of improving scpQCA is particularly inspired by a classic problem in combinatorial mathematics, the set covering problem. By treating consistency and coverage as optimization parameters, QCA problems can be transformed into scp optimization problems, thereby enhancing solution coverage.

Initially, 6 Boolean factors were established, and a complete truth table was generated based on pathways $ab+CD+ace+BDF\rightarrow OUTCOME$. Subsequently, 200 samples were randomly drawn with replacement from the complete truth table, and a varying number of samples were subjected to confounding operations on the OUTCOME values. Table \ref{tb3} presents the solutions and corresponding consistency and coverage metrics for the scpQCA, CNA, and QCApro algorithms.

\begin{table}
  \caption{High coverage experiments on scpQCA, CNA and QCApro\label{tb3}}
  \centering
  \begin{tabular}{cc>{\centering\arraybackslash}p{3.5cm}>{\centering\arraybackslash}p{3.5cm}>{\centering\arraybackslash}p{5cm}}
    \toprule
    Num.\textsuperscript{1} & Result & scpQCA & CNA & QCApro \\ 
    \midrule
    \multirow{3}{*}{0} & Sol. & $ab+CD+ace +BDF$ & $ab+ CD+ace$ & $ab+CD+ace+BDF$ \\ \cline{2-5} 
        & Con. & 1.0 & 1.0 & 1.0 \\ \cline{2-5} 
        & Cov. & 1.0 & 0.909 & 1.0 \\ \midrule
    \multirow{3}{*}{1} & Sol. & $ab+CD+BDF$ & $ab+CD+BDF$ & $ab+ CD+ace+BDF$ \\ \cline{2-5} 
       & Con. & 1.0 & 1.0 & 1.0 \\ \cline{2-5} 
       & Cov. & 0.981 & 0.981 & 0.9903 \\ \midrule
    \multirow{3}{*}{3} & Sol. & $ab+CD+ace +BDF$ & $ab+ CD+BDF$ & $ab+ACD+aDe+BDF$ \\ \cline{2-5} 
        & Con. & 0.973 & 0.99 & 0.988 \\ \cline{2-5} 
        & Cov. & 1.0 & 0.917 & 0.769 \\ \midrule
    \multirow{3}{*}{5} & Sol. & $ab+CD+ace +BDF$ & $ab+ CD+ace +BDF$ & $ab+CD+BDF+acde$ \\ \cline{2-5} 
        & Con. & 0.990 & 0.990 & 0.989 \\ \cline{2-5} 
        & Cov. & 0.960 & 0.960 & 0.899 \\ \midrule
    \multirow{3}{*}{10} & Sol. & $ab+CD+ace +BDF$ & $ab+CD+BDF$ & $ab+aCD+aCE+bCD+BDF +acdef$ \\ \cline{2-5} 
        & Con. & 0.965 & 0.981 & 1.0 \\ \cline{2-5} 
        & Cov. & 0.949 & 0.906 & 0.880 \\ \midrule
    \multirow{3}{*}{15} & Sol. & $ab+CD+BDF +aef$ & $ab+aD+CD$ & $ACD+ace+BDEF+(CDf+aBDF +abeF+abEf +acDF)$ \\ \cline{2-5} 
        & Con. & 0.907 & 0.918 & 0.948 \\ \cline{2-5} 
        & Cov. & 0.924 & 0.848 & 0.876 \\ \midrule
    \multirow{3}{*}{20} & Sol. & $ab+CD+BDF$ & $ab+CD+BDF$ & $abf+aDe+CDe+BDEF+ABCEf +(abCd+ abcE + aCDf+ BcdeF)$ \\ \cline{2-5} 
        & Con. & 0.891 & 0.891 & 0.95 \\ \cline{2-5} 
        & Cov. & 0.882 & 0.882 & 0.775 \\ 
  \bottomrule
  \end{tabular}
   \begin{tablenotes}
     \item[1] \textsuperscript{1}Confounding numbers.
   \end{tablenotes}
\end{table}

\subsection{Multi-factor}
The emergence of multi-factor problems stems from the adherence to the QMC algorithm within the QCA method. In the past, the need for multi-factor analysis in QCA was obscured by methods such as `variable merging' \cite{rutten_openness_2019} and `two-step QCA' \cite{schneider_reducing_2006}. Due to technical limitations, users have always had to adapt by employing data pre-processing or innovative approaches to work with the functionally limited qmcQCA.

Due to the increasing demand for factors in csQCA and the widespread application of mvQCA, multi-factor analysis is an urgent issue that needs to be addressed. The challenge for QMC in dealing with multi-factor problems lies in the fact that the number of variables can significantly impact its computational memory requirements. Normal computers often do not have the capacity to accommodate the exponentially growing spatial demands, leading to frequent `out of memory' errors or the program stalling without producing results. scp algorithm \cite{slavik_tight_1997}, on the other hand, trades time for space by employing a serial traversal method to screen all candidate rules (atomic minus-formulas), ensuring that a greater number of users can conveniently utilize QCA.

Similarly, generate a 20-factor Boolean truth table based on pathways $ab+CD+ace+BDF\rightarrow OUTCOME$ at first. Then gradually grow the number of factors input and evaluate the three methods' performance in Table \ref{tb4}.

\begin{table}
  \caption{Multi-factor experiments on scpQCA, CNA and QCApro\label{tb4}}
  \centering
  \begin{tabular}{cc>{\centering\arraybackslash}p{3.5cm}>{\centering\arraybackslash}p{3.5cm}>{\centering\arraybackslash}p{3.5cm}}
    \toprule
    Num.\textsuperscript{1} & Result & scpQCA & CNA & QCApro \\ 
    \midrule
    \multirow{3}{*}{6} & Sol. & $ab+CD+ace+DF$ & $ab+ CD+ace$ & $ab+ CD+ace+BDF$ \\ \cline{2-5} 
        & Con. & 0.966 & 1.0 & 1.0 \\ \cline{2-5} 
        & Cov. & 1.0 & 0.955 & 1.0 \\ \midrule
    \multirow{3}{*}{8} & Sol. & $ab+CD+ace+DF$ & $ab+ CD+ace$ &  \\ \cline{2-4} 
       & Con. & 0.966 & 1.0 & - \\ \cline{2-4} 
       & Cov. & 1.0 & 0.955 &  \\ \midrule
    \multirow{3}{*}{10} & Sol. & $ab+CD+ace+DF$ & $ab+ CD+ace$ &  \\ \cline{2-4} 
        & Con. & 0.966 & 1.0 & - \\ \cline{2-4} 
        & Cov. & 1.0 & 0.955 &  \\ \midrule
    \multirow{3}{*}{15} & Sol. & $ab+CD+ace+DF$ &  &  \\ \cline{2-3} 
        & Con. & 0.966 & - & - \\ \cline{2-3}
        & Cov. & 1.0 &  &  \\ \midrule
    \multirow{3}{*}{20} & Sol. & $ab+CD+ace+DF$ &  &  \\ \cline{2-3} 
        & Con. & 0.966 & - & - \\ \cline{2-3} 
        & Cov. & 1.0 &  &  \\ 
    \bottomrule
  \end{tabular}
   \begin{tablenotes}
     \item[1] \textsuperscript{1}Factor numbers.
   \end{tablenotes}
\end{table}

\subsection{Multi-value}
Based on our normal experiments, when the number of factors or the number of variables per factor in mvQCA increases, some QCA methods struggle to operate and the others fail to preserve the original multi-value characteristics of the results. The root cause of these issues is still fundamentally due to the use of the QMC algorithm. The QMC algorithm was initially designed for Boolean simplification \cite{abdalla_introducing_2015} and is not equipped to handle multi-value data directly. To compensate for this deficiency, a common approach in the past for mvQCA was to split a multi-value variable into multiple binary variables for processing \cite{ragin_comparative_2014}, which made it difficult to revert cases back to their original values and increase the difficult on multi-factor.

In fact, QCA's utilization of counterfactuals and multi-value variables (Jordan 2011) is something that most statistical analysis methods overlook, and it is one of the greatest strengths of QCA. For the existence of mvQCA, QCA can deal with the combined impact of variables that represent degrees, categories, or relational scenarios with other factors. scpQCA treats the multi-value nature of mvQCA data as `one-hot encoding', freeing researchers from the limitations of high multi-value issues and allowing for greater dissemination of mvQCA.

In this experiment, Table \ref{tb5} generate the test dataset with 5 factors for three times, each has different value number and the configuration pathways.

\begin{table}
  \caption{Multi-value experiments on scpQCA, CNA and QCApro\label{tb5}}
  \centering
  \begin{tabular}{>{\centering\arraybackslash}p{4cm}c>{\centering\arraybackslash}p{3cm}>{\centering\arraybackslash}p{3cm}c}
    \toprule
    Variable classes & Result metrics & scpQCA & CNA & QCApro \\ 
    \midrule
    3-value: $A0*B0+B1*C1+C2 *D2+D0*E0 =OUTCOME$  & Sol. & $A0*B0+B1*C1+D0*E0+A1*E1+C2*D2$ & $A0*B0+ A1*E1+ B1*C1+ D0*E0$ &  \multirow{3}{*}{Out of memory}\\ \cline{2-4}
    & Con. & 1.0 & 1.0 &  \\ \cline{2-4}
    & Cov. & 1.0 & 0.906 &  \\ 
    \midrule
    4-value: $A0*B0+B1*C1 +C2*D2+D3*E3 +E0*A0+A1*B1 +B2*C2+C3*D3 +D0*E0 =OUTCOME$ & Sol. & $D0*E0+C3*D3+B1*C1+A0*E0+A1*B1+D3*E3+A0*B0$ & $A0+ A1*B1+ C3*D3+ D0*E0$ & \multirow{3}{*}{Out of memory} \\ \cline{2-4}
    & Con. & 1.0 & 0.914 & \\ \cline{2-4}
    & Cov. & 0.9048 & 0.762 & \\ 
    \midrule
    5-value: $A4+A0*B0+B1 *C1+C2*D2+D3 *E3+E0*A0+A1 * B1+B 2* C 2+C 3 *D3+DO*EO =OUTCOME$ & Sol. & $A4+A1* B1+C3* D3+B2* C2+C2*D2+B1*C1+D0*E0$ & $A4+ A1*B1+ B1*C1+ B2*C2$ & \multirow{3}{*}{Out of memory}\\ \cline{2-4}
    & Con. & 1.0 & 1.0 & \\ \cline{2-4}
    & Cov. & 0.96 & 0.76 & \\ 
    \bottomrule
  \end{tabular}
\end{table}

\section{Comparison Cases{\label{sc3}}}
This section compares the similarities and differences between scpQCA and the QCA methods from past literature across various types of academic datasets, highlighting the improvements and advantages of scpQCA in terms of case coverage, handling of multiple variables, and the ability to accommodate multiple values per variable.

\subsection{Openness and Innovation}
The dataset from \cite{rutten_openness_2019} encompasses 5 factors and 108 cases in fuzzy type. Rutten analysed the innovation conditions across 108 regions in Northwest Europe and identified several underlying factors as variables to examine the causal relationships between these factors and regional innovation. The 108 events represent the 108 regions in Northwest Europe. The five factors represent the following: `analytical knowledge' (scientific and high-tech knowledge), `synthetic knowledge' (applied and engineering knowledge), `economic diversity' (diversified economic activities related to knowledge creation), `cultural melting pot' (attitudes towards ethnic and religious minorities, gender equality, and homosexuality), and `knowledge self-expression' (social values of exchanging knowledge and ideas). After indirect calibration, 19 cases (19 regions) were categorized as regions with less innovation ($reginnovation=0$), and 89 cases as regions with more innovation ($reginnovation=1$).

Rutten set the consistency threshold of the truth table at 0.85 and the minimum event coverage threshold at 5. scpQCA method preprocesses the data in identical criterion. The truth table and candidate rule table for the `Openness and Innovation' dataset are shown in the aforementioned Table \ref{tb1} and \ref{tb2}. Table \ref{tb6} compares the results obtained from the `Openness and Innovation' dataset by both qmcQCA and scpQCA, respectively.

\begin{table}
  \caption{The comparison between qmcQCA and scpQCA on Rutten's (2019) dataset\label{tb6}}
  \centering
  \begin{tabular}{cccccccc}
    \toprule
     & \multicolumn{4}{c}{qmcQCA} & \multicolumn{3}{c}{scpQCA} \\
    \cmidrule(lr){2-5} \cmidrule(lr){6-8}
    Configuration      & I & II & III & IV & 1 & 2 & 3 \\
    \midrule
    Analytical knowledge &  & $\bullet$ & $\bullet$ & $\bullet$ &  & $\bullet$ & $\bullet$ \\
    Synthetic knowledge &  & $\bullet$ & $\bullet$ &  &  & $\bullet$ &  \\
    Economic diversity & $\bullet$ & $\bullet$ &  & $\bullet$ & $\bullet$ &  &  \\
    Melting pot & $\bullet$ &  & $\bullet$ &  & $\bullet$ &  &  \\
    Self-expression &  &  &  & $\bullet$ &  &  & $\bullet$ \\
    \midrule
    Solution coverage & \multicolumn{4}{c}{0.7895} & \multicolumn{3}{c}{0.8315}  \\
    Solution consistency & \multicolumn{4}{c}{0.9441} & \multicolumn{3}{c}{0.8506} \\
    Consistency threshold & \multicolumn{4}{c}{0.85} & \multicolumn{3}{c}{0.8} \\
    Frequency threshold & \multicolumn{4}{c}{5} & \multicolumn{3}{c}{5} \\
    Truth table rows/Candidate rules & \multicolumn{4}{c}{9 rows} & \multicolumn{3}{c}{14 rules} \\
    \bottomrule
  \end{tabular}
  \begin{tablenotes}
    \item 1. The shaded factors are the necessary conditions identified by scpQCA. Solid circles indicate that the factor has an effect and the effect is positive, while hollow circles signify the presence of the factor's influence and the effect is negative
  \end{tablenotes}
\end{table}

Configuration-I obtained from qmcQCA. Additionally, Configuration-2 from scpQCA is a superset of Configuration-II and Configuration-III derived from qmcQCA. In other words, Configuration-2 from scpQCA encompasses both Configuration-II and Configuration-III from the qmcQCA results. Lastly, Configuration-3 from scpQCA is a superset of Configuration-IV from qmcQCA, with the distinction that Configuration-3 simplifies the `Economic diversity' factor from Configuration-IV.

\subsection{Democracy Distribution}
The Haesebrouck (2019) dataset comprises 22 cases and 8 factors with fuzzy data \cite{haesebrouck_alternative_2019}. The study investigates the potential reasons behind the varying financial contribution ratios of democratic countries when they pledged to support the United Nations Interim Force in Lebanon (UNIFIL) in 2006 \cite{haesebrouck_democratic_2015}. After calibrating the raw fuzzy data using an indirect calibration method, the 22 democratic countries were categorized into two groups: those with minimal contributions ($LC=0$) and those with significant contributions ($LC=1$). The 8 factors in the dataset are: Military Capability (MC), Military Stretchability (MS), prior Involvement in Peacekeeping activities (PI), high Geographic Proximity (GP), whether the Executive branch is Left-leaning (LE), whether the Parliamentary attitude is Left-leaning (LP), the proximity of the next national Election Date (ED), and whether the Parliament has Voting rights (PV).

Consistent with the data processing methods of Schneider (2019) \cite{schneider_two-step_2019} and Haesebrouck (2019) \cite{haesebrouck_alternative_2019}, scpQCA sets the $consistency$ threshold at 0.8 and the $unique\_cover$ threshold at 2 before conducting the analysis. Table \ref{tb7} displays the final results calculated by the two versions of `two-step QCA' and scpQCA, with the `two-step QCA' results derived from Schneider (2019) and Haesebrouck (2019), respectively.

\begin{table}
  \caption{The comparison between two-step QCA methods and scpQCA on Haesebrouck's (2015) dataset\label{tb7}}
  \centering
  \begin{tabular}{c>{\centering\arraybackslash}p{1cm}>{\centering\arraybackslash}p{1cm}>{\centering\arraybackslash}p{1cm}>{\centering\arraybackslash}p{1cm}>{\centering\arraybackslash}p{1cm}>{\centering\arraybackslash}p{1cm}>{\centering\arraybackslash}p{1cm}>{\centering\arraybackslash}p{1cm}>{\centering\arraybackslash}p{1cm}}
    \toprule
    & \multicolumn{3}{c}{Schneider2019} & \multicolumn{3}{c}{Haesebrouck2019} & \multicolumn{3}{c}{scpQCA} \\
    \cmidrule(lr){2-4} \cmidrule(lr){5-7} \cmidrule(lr){8-10}
    Configuration & I & II & III & A & B & C & 1 & 2 & 3 \\
    \midrule
    MC & $\bullet$ & $\bullet$ &  & $\bullet$ & $\bullet$ &  & $\bullet$ & $\bullet$ &  \\
    MS &  &  &  &  & $\circ$ & $\circ$ &  & $\circ$ & $\circ$  \\
    PI &  &  & $\bullet$ &  &  & $\bullet$ &  &  & $\bullet$  \\
    GP &  &  &  &  &  &  &  &  &  \\
    LE & $\bullet$ & $\bullet$ & $\bullet$ & $*$ & $\bullet$ & $*$ & $*$ & $*$ & $*$ \\
    LP & $\bullet$ &  &  & $\bullet$ &  & $\bullet$ & $\bullet$ &  & $\bullet$ \\
    ED & $\bullet$ & $\bullet$ & $\bullet$ & $*$ & $*$ & $*$ & $*$ & $*$ & $*$ \\
    PV &  & $\circ$ & $\bullet$ &  &  &  &  & $\circ$ &  \\
    \midrule
    Covered cases & BE, ES, FR, GR, PL, PT & BE, FR, GB, GR, IT, PL, PT & CZ, FI, IE, SK & BE, ES, FR, GR, PL, PT & BE, ES, FR, GR, IT, PL, PT & FI, FR, PL, PT & BE, ES, FR, GR, PL, PT & BE, FR, GR, IT, PL, PT & FI, FR, PL, PT \\
    Consistency & 1.0 & 0.8571 & 0.5 & 1.0 & 1.0 & 1.0 & 1.0 & 1.0 & 1.0 \\
    Unique coverage & 1 & 2 & 4 & 0 & 1 & 1 & 1 & 1 & 1 \\
    \midrule
    Solution coverage & \multicolumn{3}{c}{1.0} & \multicolumn{3}{c}{0.8888} & \multicolumn{3}{c}{0.8888} \\
    Solution consistency & \multicolumn{3}{c}{0.8182} & \multicolumn{3}{c}{1.0} & \multicolumn{3}{c}{1.0} \\
    \bottomrule
  \end{tabular}
  \begin{tablenotes}
    \item 1. The $*$ factors are the necessary conditions. Solid circles indicate that the factor has an effect and the effect is positive, while hollow circles signify the presence of the factor's influence and the effect is negative
    \item 2. scpQCA has recalculated the consistency, coverage, and overall solution consistency and coverage for the rules established by Schneider (2019) and Haesebrouck (2019).
  \end{tablenotes}
\end{table}

The results generated by scpQCA are nearly identical to those obtained by Schneider (2019) and Haesebrouck (2019) using the more cumbersome two-step QCA approach, while maintaining higher consistency and coverage. Both Schneider (2019) and Haesebrouck (2019) prioritized the analysis and screening of these four factors (MS, MC, PI, and GP) as distant factors before proceeding to the second phase of the two-step QCA; in contrast, scpQCA can simultaneously analyze all factors in a single computation. Consequently, scpQCA has eliminated the traditional weaknesses of the qmcQCA algorithm in handling multi-factor analyses and offers a more refined functionality.

\subsection{Variations in Party Bans}
The Pban dataset primarily investigates the main reasons behind the emergence and enforcement of political party bans in Sub-Saharan Africa. It includes 48 cases with five multi-value characteristic variables, which are: Colonial background C (“2” represents British, “1” represents French, “0” represents other); Type of pre-regime competition F (“2” indicates none, “1” indicates a limited number, “0” indicates multiple regimes); Transition mode before 1990 T (“2” represents managerial, “1” represents negotiated, “0” represents democratic); and Ethnic conflict V (“1” indicates present, “0” indicates absent). The output variable is the introduction of party ban provisions (PB), where “1” indicates presence and “0” indicates absence.

The Pban data originates from the CNA algorithm package, and the results from the mvQCA application in CNA are not entirely consistent with the original data source (Hartmann and Kemmerzell 2010). However, when using the same input parameter constraints, the results from CNA, QCApro, and scpQCA are almost completely identical and the solutions from these three methods are compared in Table \ref{tb8}.

\begin{table}
  \caption{The comparison between CNA, QCApro and scpQCA on Pban dataset\label{tb8}}
  \centering
  \begin{tabular}{c>{\centering\arraybackslash}p{0.7cm}>{\centering\arraybackslash}p{0.7cm}>{\centering\arraybackslash}p{0.7cm}>{\centering\arraybackslash}p{0.7cm}>{\centering\arraybackslash}p{0.7cm}>{\centering\arraybackslash}p{0.7cm}>{\centering\arraybackslash}p{0.7cm}>{\centering\arraybackslash}p{0.7cm}>{\centering\arraybackslash}p{0.7cm}>{\centering\arraybackslash}p{0.7cm}>{\centering\arraybackslash}p{0.7cm}>{\centering\arraybackslash}p{0.7cm}}
    \toprule
    & \multicolumn{4}{c}{CNA} & \multicolumn{4}{c}{QCApro} & \multicolumn{4}{c}{scpQCA} \\
    \cmidrule(lr){2-5} \cmidrule(lr){6-9} \cmidrule(l){10-13}
    Configuration & I & II & III & IV & A & B & C & D & 1 & 2 & 3 & 4 \\
    \midrule
    C & 1 &  & 0 & 2 & 1 &  & 0 & 2 & 1 &  &  & 2 \\
    F &  & 2 & 1 &  &  & 2 &  &  &  & 2 &  &  \\
    T &  &  &  &  &  &  & 2 &  &  &  & 2 &  \\
    V &  &  &  & 0 &  &  &  & 0 &  &  &  & 0 \\
    \midrule
    Con. & 1.0 & 1.0 & 1.0 & 1.0 & 1.0 & 1.0 & 1.0 & 1.0 & 1.0 & 1.0 & 0.94 & 1.0 \\
    Cov. & 17 & 26 & 3 & 7 & 17 & 26 & 12 & 7 & 17 & 26 & 34 & 7 \\
    Unique coverage & 7 & 13 & 3 & 4 & 7 & 4 & 3 & 4 & 3 & 2 & 7 & 2 \\
    \midrule
    Sol. cov. & \multicolumn{4}{c}{0.952} & \multicolumn{4}{c}{0.952} & \multicolumn{4}{c}{1.0} \\
    Sol. con. & \multicolumn{4}{c}{1.0} & \multicolumn{4}{c}{1.0} & \multicolumn{4}{c}{0.9545} \\
    \bottomrule
  \end{tabular}
\end{table}

Table \ref{tb8} reveals that configuration-I, A, and 1 are identical; configuration-II, B, and 2 are identical; and configuration-IV, D, and 4 are identical. The differences in the third configurations of the three solutions lead to variations between solution coverage and consistency, which is reflected in scpQCA method's higher coverage and slightly lower consistency. All three methods can directly compute mvQCA and achieve very similar solutions, indicating that scpQCA maintains the stability inherent in QCA methodology. 

\section{Robustness{\label{sc4}}}
This section assesses the internal and external robustness of scpQCA using the definitions and methods described in \cite{rutten_applying_2022}. Internal robustness is determined by running the program with various consistency thresholds or frequency thresholds and comparing the final outcomes. External robustness is evaluated by extracting a set number of events from the dataset to create different computational case datasets, which are then used to test and compare the results. This robustness research is comparable to another QCA robustness study conducted by \cite{oana_robustness_2024}. Integrating the robustness definitions from both, here we focus on comparing the internal and external robustness on Openness and Innovation and Democracy Distribution case datasets.

\subsection{Internal validity}
Internal validity serves to test whether scpQCA yields significantly different results under varying consistency and frequency threshold values, which are the input parameters. The Openness and Innovation dataset is used to test internal validity by adjusting the consistency threshold, while the Democracy Distribution dataset employs frequency threshold adjustments.

First, run scpQCA on the Openness and Innovation dataset and set the consistency threshold to 0.80, 0.75, and 0.7 to calculate the solution configurations, as shown in Table \ref{tb9}. It can be intuitively found that when the consistency threshold drops to 0.75, a new rule 4 `Analytical knowledge*Economic diversity' appears in the configuration. This combination is a superset of the configurations II and IV in Table \ref{tb6}. At the same time, the coverage of $sufficient=0.75$ and $0.70$ significantly increases, which is due to the lower consistency threshold.

\begin{table}
  \caption{Internal validity for adjustive sufficient consistency thresholds on Rutten's (2019) Dataset}
  \label{tb9}
  \centering
  \begin{tabular}{c>{\centering\arraybackslash}p{0.7cm}>{\centering\arraybackslash}p{0.7cm}>{\centering\arraybackslash}p{0.7cm}>{\centering\arraybackslash}p{0.7cm}>{\centering\arraybackslash}p{0.7cm}>{\centering\arraybackslash}p{0.7cm}>{\centering\arraybackslash}p{0.7cm}>{\centering\arraybackslash}p{0.7cm}>{\centering\arraybackslash}p{0.7cm}>{\centering\arraybackslash}p{0.7cm}>{\centering\arraybackslash}p{0.7cm}}
    \toprule
    & \multicolumn{3}{c}{sufficient=0.80} & \multicolumn{4}{c}{sufficient=0.75} & \multicolumn{4}{c}{sufficient=0.70} \\
    \cmidrule(lr){2-4} \cmidrule(lr){5-8} \cmidrule(l){9-12}
    Configuration & 1 & 2 & 3 & 1 & 2 & 3 & 4 & 1 & 2 & 3 & 4 \\
    \midrule
    Analytical knowledge &  & $\bullet$ & $\bullet$ &  & $\bullet$ & $\bullet$ & $\bullet$ &  & $\bullet$ & $\bullet$ & $\bullet$ \\
    Synthetic knowledge &  & $\bullet$ &  &  & $\bullet$ &  &  &  & $\bullet$ &  & \\
    Economic diversity & $\bullet$ &  &  & $\bullet$ &  &  & $\bullet$ & $\bullet$ &  &  & $\bullet$  \\
    Melting pot & $\bullet$ &  &  & $\bullet$ &  &  &  & $\bullet$ &  &  & \\
    Self-expression &  &  & $\bullet$ &  &  & $\bullet$ &  &  &  & $\bullet$ &   \\
    \midrule
    Solution coverage & \multicolumn{3}{c}{0.8315} & \multicolumn{4}{c}{0.8652} & \multicolumn{4}{c}{0.8652}\\
    Solution consistency & \multicolumn{3}{c}{0.8506} & \multicolumn{4}{c}{0.8462} & \multicolumn{4}{c}{0.8462}\\
    Consistency threshold & \multicolumn{3}{c}{0.80} & \multicolumn{4}{c}{0.75} & \multicolumn{4}{c}{0.70}\\
    Frequency threshold & \multicolumn{3}{c}{5} & \multicolumn{4}{c}{5} & \multicolumn{4}{c}{5}\\
    Num. of Candidate rules & \multicolumn{3}{c}{14 rules} & \multicolumn{4}{c}{19 rules} & \multicolumn{4}{c}{25 rules}\\
    \bottomrule
  \end{tabular}
\end{table}

Next, keep the consistency threshold constant at 0.8, and set the frequency threshold to 2, 3, 4, and 5. Run scpQCA on the multi-factor dataset, and display the results in Table \ref{tb10}. It can be seen from Table 10 that as the frequency threshold gradually increases, the number of candidate rules decreases. In $frequency=2$, $3$, $4$, and $5$, the configurations are basically the same, except that when the frequency threshold is 5, the solution of $frequency=5$ omits configuration 3.

\begin{table}
  \caption{Internal validity for adjustive frequency thresholds on Haesebrouck's (2015) Dataset}
  \label{tb10}
  \centering
  \begin{tabular}{c>{\centering\arraybackslash}p{0.7cm}>{\centering\arraybackslash}p{0.7cm}>{\centering\arraybackslash}p{0.7cm}>{\centering\arraybackslash}p{0.7cm}>{\centering\arraybackslash}p{0.7cm}>{\centering\arraybackslash}p{0.7cm}>{\centering\arraybackslash}p{0.7cm}>{\centering\arraybackslash}p{0.7cm}>{\centering\arraybackslash}p{0.7cm}>{\centering\arraybackslash}p{0.7cm}>{\centering\arraybackslash}p{0.7cm}}
    \toprule
    & \multicolumn{3}{c}{frequency=2} & \multicolumn{3}{c}{frequency=3} & \multicolumn{3}{c}{frequency=4} & \multicolumn{2}{c}{frequency=5} \\
    \cmidrule(lr){2-4} \cmidrule(lr){5-7} \cmidrule(l){8-10} \cmidrule(l){11-12}
    Test & 1 & 2 & 3 & 1 & 2 & 3 & 1 & 2 & 3 & 1 & 2 \\
    \midrule
    MC & $\bullet$ & $\bullet$ &  & $\bullet$ & $\bullet$ &  & $\bullet$ & $\bullet$ &  & $\bullet$ & $\bullet$ \\
    MS &  & $\circ$ & $\circ$  &  & $\circ$ & $\circ$  &  & $\circ$ & $\circ$  &  & $\circ$  \\
    PI &  &  & $\bullet$ &  &  & $\bullet$ &  &  & $\bullet$ & & \\
    GP &  &  &  &  &  &  &  &  &  \\
    LE & $*$ & $*$ & $*$ & $*$ & $*$ & $*$ & $*$ & $*$ & $*$ & $*$ & $*$  \\
    LP & $\bullet$ &  & $\bullet$ & $\bullet$ &  & $\bullet$ & $\bullet$ &  & $\bullet$ & $\bullet$ &  \\
    ED & $*$ & $*$ & $*$ & $*$ & $*$ & $*$ & $*$ & $*$ & $*$ & $*$ & $*$ \\
    PV & & $\circ$ &  & & $\circ$ &  & & $\circ$ &  & & $\circ$ \\
    \midrule
    Solution coverage & \multicolumn{3}{c}{0.8888} & \multicolumn{3}{c}{0.8888} & \multicolumn{3}{c}{0.8888} & \multicolumn{2}{c}{0.8888} \\
    Solution consistency & \multicolumn{3}{c}{1.0} & \multicolumn{3}{c}{1.0} & \multicolumn{3}{c}{1.0} & \multicolumn{2}{c}{1.0} \\
    Consistency threshold & \multicolumn{3}{c}{0.8} & \multicolumn{3}{c}{0.8} & \multicolumn{3}{c}{0.8} & \multicolumn{2}{c}{0.8} \\
    Frequency threshold & \multicolumn{3}{c}{2} & \multicolumn{3}{c}{3} & \multicolumn{3}{c}{4} & \multicolumn{2}{c}{5} \\
    Num. of candidate rules & \multicolumn{3}{c}{59 rules} & \multicolumn{3}{c}{37 rules} & \multicolumn{3}{c}{13 rules} & \multicolumn{2}{c}{9 rules} \\
    \bottomrule
  \end{tabular}
\end{table}

\subsection{External validity}
External validity pertains to the causal accuracy of configurations in relation to the dependent variable of cases outside the training set. Therefore, the design of control variables primarily depends on case selection and scope conditions \cite{goertz_multimethod_2017}. An accepted method for testing external validity is to randomly remove a subset of cases from the original dataset and subsequently re-analyse the truth table \cite{rutten_openness_2019}.

To verify external validity, the first step is to generate the new test datasets. We repeatedly withdrew a random samples equivalent to 10\% of the total number of cases from `Openness and Innovation' and `Democracy Distribution' cases for 10 times, discarding these samples and using the remaining cases to form a new test dataset for each time. Subsequently, run scpQCA on the test datasets, and compare the solution configuration from the test dataset with the original solution configuration. The results of the external validity\footnote{The comparison results are categorized into four types: replicated, superset, subset, and not identified. These categories represent the set relationship between the cases covered by the test configurations and those covered by the original dataset configurations. Typically, an exact match is considered a replicated; when the sampled rule lacks a factor but is otherwise consistent with the original dataset, it is determined to be a superset; when an additional factor is included in the sampled rule but is otherwise consistent with the rules of the original dataset, it is determined to be a subset; not identified is challenging to define, hence the determination method used in this paper is derived from \cite{rutten_openness_2019}.} comparison are shown in Table \ref{tb11}.

\begin{table}
  \caption{External validity for Rutten's (2019) and Haesebrouck's (2015) Dataset}
  \label{tb11}
  \centering
  \begin{tabular}{ccccccccc}
    \toprule
    & \multicolumn{4}{c}{Openness and Innovation} & \multicolumn{4}{c}{Democracy Distribution} \\
    \cmidrule(lr){2-5} \cmidrule(lr){6-9}
    Configurations & 1 & 2 & 3 & Number & 1 & 2 & 3 & Number \\
    \midrule
    Replicated & 5 & 10 & 10 & 25 & 6 & 5 & 8 & 19 \\
    Superset & - & - & - & 0 & - & 4 & - & 4 \\
    Subset & - & - & - & 0 & - & - & - & 0 \\
    not Identified & - & - & - & 9 & - & - & - & 2 \\
    \midrule
    Accuracy(except not identified) & 0.5 & 1.0 & 1.0 & 0.7353 & 0.6 & 0.9 & 0.8 & 0.92 \\
    \bottomrule
  \end{tabular}
\end{table}

The `Openness and Innovation' dataset and the `Democracy Distribution' dataset each have a total of 34 and 25 configurations, respectively. Among them, the `Openness and Innovation' contains two not identified configurations, while the `Democracy Distribution' also includes two not identified configurations. The Number indicator in the horizontal direction refers to the count of how many times the particular category of comparison result appears across 10 random tests. The vertical Accuracy metric refers to the frequency with which configurations replicated in the entire dataset across 10 repeated tests, with a higher frequency indicating greater robustness of the configuration. The results from Table \ref{tb11} indicate that scpQCA possesses considerable external validity.

\section{Discussion{\label{sc5}}}
In summary, scpQCA addresses three issues present in the handling of mvQCA: low coverage, limited number of factors that can be processed, and limited number of variable values. Moreover, the analytical steps of scpQCA are almost identical to those of existing QCA methods, utilizing the Python programming language and following the two steps: 
\begin{enumerate}
  \item As with previous QCA software, first calculate the necessary conditions; if there are factors that meet the necessary conditions, exclud them in the subsequent calculation of sufficient conditions;
  \item Based on the factors that have excluded the necessary conditions, calculate the sufficient configurations.
\end{enumerate}

In addition to being highly effective in addressing mvQCA problems, any data that meet the computational requirements can yield results with both considerable consistency and coverage by scpQCA. Such data must satisfy the properties of set theory, such as the certainty of cases, the uniqueness of cases, and the lack of (temporal) order among cases. Furthermore, scpQCA is not sensitive to uneven sample distribution, but it is still recommended to pre-process event data by removing duplicates before computation; otherwise, the final results may implicitly contain the information that high-frequency events have higher weight. Beyond temporal events, scpQCA has a wide range of applications. We have also provided a set of operational parameters that may be necessary for users. 

However, the actual limits of scpQCA remain unknown. Although it can generate results that involve more variables, cover a broader range of variable values, and are more concise with higher consistency than the original QCAs, these results, derived directly from computation, serve as a reference for causal inference. It is still essential for the operator to make judgments carefully based on theoretical knowledge.

Finally, we would like to emphasize three key points in conclusion. First, we believe that scpQCA analysis is the very initial step, which should be followed by process tracing to further support the rationality of the scpQCA solutions. Second, in specific application domains, scpQCA is more appropriate and useful for handling data on national crisis behaviors, which are primarily categorical and ordinal in nature. Lastly, scpQCA may possess certain predictive inference capabilities. Our preliminary results indicate that the solution paths obtained from scpQCA are similar to those derived from statistical machine learning models such as decision tree. However, it should be noted that as a qualitative-quantitative comparative method, the accuracy of scpQCA results depends on the completeness of the cases and the accuracy of the data calibration. Therefore, scpQCA is more of a case-centered method than a theory-centered method \cite{rohlfing_check_2021}, and it requires continuous improvement in data quality to ensure the correctness and stability of the results.

\bibliographystyle{unsrt}  
\bibliography{references}

\end{document}